\documentclass[aps,prapplied,preprint,superscriptaddress,nofootinbib]{revtex4-2}

\usepackage{amsmath,amssymb,bm}
\usepackage{graphicx}
\usepackage{subcaption} 
\usepackage{caption} 
\usepackage[colorlinks=true,linkcolor=blue,citecolor=blue,urlcolor=blue]{hyperref}

\begin{document}

\title{Comment on ``Fundamental limit of phonon Tesla valve for heat rectification from first principles''}

\begin{abstract}
Thermal rectification is a two-terminal property: the same device must carry different heat-current magnitudes when two reservoir temperatures are interchanged. In this Comment on Ref.~\cite{WuHu2026}, we ask whether the reported ratio $\gamma=R_{\mathrm{f}}/R_{\mathrm{b}}$ can represent such a two-terminal rectification ratio within the fixed-background linearized phonon Boltzmann transport equation (BTE) used in that work. We show that, for a passive two-terminal reservoir problem, any fixed linearized BTE whose scattering operator preserves a uniform equilibrium temperature shift, and whose boundary-value problem is well posed, gives equal forward and reverse resistance magnitudes. A direction-dependent response can still occur under a prescribed gradient-weighted source--sink protocol, but such a response should be distinguished from two-terminal thermal rectification.
\end{abstract}
\author{Samuel Huberman}
\email{samuel.huberman@mcgill.ca}
\affiliation{Department of Chemical Engineering, McGill University, Montreal, Canada}
\affiliation{Department of Physics, McGill University, Montreal, Canada}

\author{Aleksei Sokolov}

\affiliation{Institute of Mechanics, Technische Universität Berlin, Berlin, Germany}
\affiliation{Research Center for Microperipheric Technologies, Technische Universität Berlin, Berlin, Germany}

\date{\today}

\maketitle

\newpage
Thermal rectification is commonly defined by comparing the same two-terminal device under interchanged reservoir temperatures: the heat-current magnitudes associated with $Q(T_1,T_2)$ and $Q(T_2,T_1)$ must differ for $T_1\neq T_2$~\cite{Budaev2017}. This distinction is important in nonlocal phonon transport, where different source and sink prescriptions can produce different internal temperature fields even within a linear theory. In this Comment, we distinguish these notions for the phonon Tesla valve of Ref.~\cite{WuHu2026}. We do not question that asymmetric geometry can yield direction-dependent nonequilibrium phonon trajectories under a prescribed injection protocol. Rather, we show that interchanged thermal reservoirs give equal forward and reverse resistances in the fixed-background linearized BTE, whereas a gradient-driven source--sink protocol need not obey the same symmetry.

Ref.~\cite{WuHu2026} writes the linearized deviational-energy BTE as
\begin{equation}\label{eq:lbte}
\frac{\partial \delta f_\lambda}{\partial t}
+\mathbf v_\lambda\cdot \nabla \delta f_\lambda
=\sum_{\lambda'} A_{\lambda,\lambda'}\delta f_{\lambda'},
\end{equation}
where $\delta f_\lambda=\hbar\omega_\lambda\delta n_\lambda$. The modal specific heat is defined as
\begin{equation}
C_\lambda \equiv \hbar\omega_\lambda
\frac{\partial n^0_\lambda}{\partial T}
\end{equation}
up to a normalizing factor consisting of the unit cell volume and number of $\mathbf q$ points in the discretized Brillouin zone. One can then define a small local-equilibrium temperature shift $\Theta$ which is much less than the global background temperature $T_0$ such that
\begin{equation}
\delta f^{\rm eq}_\lambda=C_\lambda\Theta .
\end{equation}
Because a spatially uniform shift of the Bose-Einstein distribution is an equilibrium state, the collision operator must satisfy
\begin{equation}
\sum_{\lambda'}A_{\lambda,\lambda'}C_{\lambda'}=0,
\end{equation}
and the heat flux associated with the uniform equilibrium perturbation vanishes:
\begin{equation}
\mathbf Q^{\rm eq}=\sum_\lambda \mathbf v_\lambda C_\lambda\Theta=0,
\end{equation}
by time-reversal symmetry of phonon modes. Equivalently, for every mode $\lambda=(\mathbf q,s)$ there is a time-reversed partner $\bar\lambda=(-\mathbf q,s)$ with $C_{\bar\lambda}=C_\lambda$ and $\mathbf v_{\bar\lambda}=-\mathbf v_\lambda$.

It is useful to introduce the linear transport operator associated with Eq.~(1):
\begin{equation}
\mathcal L[X]_\lambda
\equiv
\frac{\partial X_\lambda}{\partial t}
+\mathbf v_\lambda\cdot\nabla X_\lambda
-\sum_{\lambda'}A_{\lambda,\lambda'}X_{\lambda'} .
\end{equation}
Equation~(1) is therefore the statement that the relevant deviational-energy distribution is in the null space of this operator, \(\mathcal L[\delta f]=0\). The operator \(\mathcal L\) is linear in its argument as long as the scattering matrix and velocities are evaluated about the same fixed background temperature. For solutions $X_\lambda$, the terminal heat current is the linear functional

\begin{equation}\label{eq:heatcurrent}
Q_i[X]=\int_{\Gamma_i}\sum_\lambda
(\mathbf v_\lambda\cdot\mathbf n_i)X_\lambda(\mathbf r)\,dS,
\end{equation}

where \(\Gamma_i\) denote the cross-sectional boundary surface at terminal \(i\), where the device is connected to reservoir \(i\), and let \(\mathbf n_i(\mathbf r)\) be the unit normal on \(\Gamma_i\) pointing outward from the device into the reservoir. Now consider the valve geometry of Ref.~\cite{WuHu2026} with terminals $1$ and $2$. $F_\lambda(\mathbf r)$ denotes the forward-bias solution of Eq.~(1), $F_\lambda(\mathbf r)=\delta f^{\rm fwd}_\lambda(\mathbf r)$, and thus satisfies $\mathcal L[F]=0$ by construction. Similarly, $\widetilde F_\lambda$ denotes the reverse-bias solution. Let $F_\lambda(\mathbf r)$ be the steady solution when terminal $1$ is held at $T_1=T_0+\Theta$ and terminal $2$ is held at $T_2=T_0$, with all other boundaries passive and energy-conserving. The boundary condition at a thermal reservoir is linear in the imposed temperature perturbation: phonons from a reservoir at $T_0+\Theta_i$ have the local-equilibrium deviational distribution $C_\lambda\Theta_i$. Therefore $F$ satisfies the reservoir boundary conditions

\begin{equation}
F|_{1}=C_\lambda\Theta,\qquad F|_{2}=0 .
\end{equation}

The candidate reversed solution is
\begin{equation}\label{eq:candidate}
\widetilde F_\lambda(\mathbf r)=U_\lambda-F_\lambda(\mathbf r).
\end{equation}
where $U_\lambda=C_\lambda\Theta$ is the spatially uniform solution. Since the forward reservoir values are $C_\lambda\Theta$ and $0$, an affine transformation $F\mapsto U-F$ is needed to swap the reservoirs. If the forward reservoirs are $F|_{1}=C_\lambda\Theta$ and $F|_{2}=0$ and the backward reservoirs are $\widetilde F|_{1}=-C_\lambda\Theta$ and $\widetilde F|_{2}=0$, the transformation becomes $\widetilde F=-F$.

Because $\mathcal L$ is linear, the candidate also satisfies the same interior equation:
\begin{equation}
\mathcal L[\widetilde F]
=\mathcal L[U-F]
=\mathcal L[U]-\mathcal L[F]
=0 .
\end{equation}
It remains to check the reversed reservoir boundaries. The uniform solution has
\begin{align}
\widetilde F|_{1}&=U|_1-F|_1
=C_\lambda\Theta-C_\lambda\Theta=0,\\
\widetilde F|_{2}&=U|_2-F|_2
=C_\lambda\Theta-0=C_\lambda\Theta .
\end{align}
These are precisely the boundary conditions for the reversed thermal bias. Assuming the standard uniqueness of the steady linear BTE with these reservoir and passive-wall boundary conditions, the candidate in Eq.~\eqref{eq:candidate} is therefore the reverse-bias solution. Since the uniform equilibrium solution carries no heat current, inserting Eq.~\eqref{eq:candidate} into Eq.~\eqref{eq:heatcurrent} gives
\begin{equation}
Q_i[\widetilde F]=Q_i[U-F]=-Q_i[F], 
\end{equation}

and therefore $R_{\mathrm{f}}/R_{\mathrm{b}}= 1$. This conclusion is specific to a two-terminal device.  If a third terminal is joined to the geometry and acts a fixed bath, the affine transformation no longer maps the forward problem to the reversed problem.  Let \(F^{(i)}\) denote the solution with terminal \(i\) at \(T_0+\Theta\) and the other two terminals at \(T_0\).  The uniform solution satisfies \(U=F^{(1)}+F^{(2)}+F^{(3)}\).  Hence \(U-F^{(1)}=F^{(2)}+F^{(3)}\), which has terminals \(2\) and \(3\) both at \(T_0+\Theta\), and is not the reverse problem \(F^{(2)}\) with terminal \(3\) fixed at \(T_0\).

Passive non-reservoir boundaries, such as diffusive or specular side walls, also do not evade the argument. Specular reflection is a linear map of the incoming distribution. A diffuse adiabatic wall may be represented by a linear energy-conserving map that redistributes the outgoing deviational energy into the mode-dependent $C_{\lambda}$ distribution while enforcing zero net normal energy flux. Such a wall leaves the uniform equilibrium perturbation $U_\lambda=C_\lambda\Theta$ invariant. Hence, if both $F$ and $U$ satisfy the same passive wall rule, then $U-F$ satisfies it as well.

Ref.~\cite{WuHu2026} uses inlet conditions that are different from a standard isothermal diffuse reservoir. In the Monte Carlo simulation, sample particles at the inlet are initialized with a probability proportional to \(|C_\lambda\mathbf v_\lambda\cdot\nabla T|\), while the outlet is fixed at the background temperature and removes particles upon arrival. Only if the imposed gradient is normal to the inlet boundary does this sampling have the same angular flux weight as emission from a diffuse isothermal reservoir, and even then its amplitude is set by the imposed gradient or heating power rather than by an isothermal reservoir population $C_\lambda\Theta_i$. In Ref.~\cite{WuHu2026}, the direction of the imposed \(\nabla T\) relative to the local inlet normal is not specified in sufficient detail to identify this sampling rule with a standard isothermal reservoir condition.

Regarding the analogy to classical fluids discussed in Ref.~\cite{WuHu2026}, the relevant distinction is the role of inertia. In the low-Reynolds-number limit, the Navier--Stokes equations reduce to the linear Stokes equations, for which reversing the pressure difference reverses the velocity field and gives the same hydraulic resistance. Directional resistance in a classical Tesla valve arises only when inertial nonlinearities become important at finite Reynolds number~\cite{Nguyen2021AJP}. In this respect, the fixed-background linearized phonon BTE is more closely analogous to Stokes flow than to inertial fluid flow. In this regime, the  LBTE can be well approximated by the Viscous Heat Equations (VHE)~\cite{Simoncelli2020,Dragasevic2026}

\begin{align}
\label{eq:vhe}
\sum_{i,j=1}^{3} W_{0j}^{i} \sqrt{\bar{T} A^{j} C} \frac{\partial u^{j}(\boldsymbol{r}, t)}{\partial r^{i}} - \sum_{i,j=1}^{3} \kappa^{ij} \frac{\partial^{2} T(\boldsymbol{r}, t)}{\partial r^{i} \partial r^{j}} &= 0, \\
\sqrt{\frac{C A^{i}}{T_0 }} \sum_{j=1}^{3} W_{i0}^{j} \frac{\partial T(\boldsymbol{r}, t)}{\partial r^{j}} - \sum_{j,k,l=1}^{3} \mu^{ijkl} \frac{\partial^{2} u^{k}(\boldsymbol{r}, t)}{\partial r^{j} \partial r^{l}} &= -\sum_{j=1}^{3} \sqrt{A^{i} A^{j}} \, D_{U}^{ij} u^{j}(\boldsymbol{r}, t).
\end{align}

where $C$ is the specific heat, $W_{0j}^{i}$ is a velocity tensor, $T_0$ is the reference (equilibrium) temperature on which a perturbation is applied , $A^{i}$ is the specific momentum in direction $i$, $\mu^{ijkl}$ is the thermal viscosity tensor, $\kappa^{ij}$ is the thermal conductivity tensor, and $D_{U}^{ij}$ is the momentum dissipation rate. For details and tabulated values of the input parameters, see Ref.~\cite{Simoncelli2020}.

In the Tesla Valve geometry, the boundary conditions applied to the VHEs can be made to match those of the LBTE by setting $T|_1 = T_0 + \Theta$, $T|_2 = T_0$ and additionally imposing no-slip drift velocity corresponding to thermalised terminals $\boldsymbol{u}|_1=0$, $\boldsymbol{u}|_2=0$. In this case, the forward solution which satisfy
\begin{align}
    T = T_0 + \vartheta(\boldsymbol{r}), \quad \boldsymbol{u} = \boldsymbol{v} ( \boldsymbol{r})
\end{align}
Then the reverse solution is 
\begin{align}
    \label{eq:vhe_solution}
    \widetilde{T} = T_0 + \Theta - \vartheta(\boldsymbol{r}), \quad \widetilde{\boldsymbol{u}} =  - \boldsymbol{v} ( \boldsymbol{r})
\end{align}
However, if a nonzero terminal drift velocity is prescribed, e.g., \(\boldsymbol{u}|_1=\boldsymbol{u}_{\mathrm{in}}\neq\boldsymbol{0}\) and \(\boldsymbol{u}|_2=\boldsymbol{0}\), the affine transformation gives \(\widetilde{\boldsymbol{u}}|_1=-\boldsymbol{u}_{\mathrm{in}}\) and \(\widetilde{\boldsymbol{u}}|_2=\boldsymbol{0}\), rather than the boundary condition obtained by interchanging inlet and outlet with the same positive drift velocity. Only in the case where the inlet velocity equals the outlet velocity, does there exist an affine transformation which is a solution of VHE. 

We illustrate this effect by solving the VHE in a Tesla-valve geometry for graphite with (0.1\%) isotope concentration, ambient temperature ($T_0=80 ~\mathrm{K}$), and imposed temperature difference ($\Theta=1~\mathrm{K}$). The thermal boundary conditions are fixed at ($T_0+\Theta$) on the inlet and ($T_0$) on the outlet. At the heated inlet, we additionally impose a drift velocity normal to the boundary, ($u_{\perp}=U$), with zero tangential component, ($u_{\parallel}=0$), while the drift velocity at the outlet is kept zero. Example temperature and velocity profiles are shown in Figure 1. We sweep $U$ from 0 to 0.05 $\mu\mathrm{m/ns}$ and compute the resulting heat flux for two cases: imposing the nonzero drift velocity at the left inlet and at the right inlet, respectively, as shown in Figure 2 (a). Setting the inlet and outlet velocities to be equal removes any rectification, as shown in Figure 2 (b).

\begin{figure}[htbp]
    \centering
    \begin{subfigure}{0.48\textwidth}
        \centering
        \includegraphics[width=\textwidth]{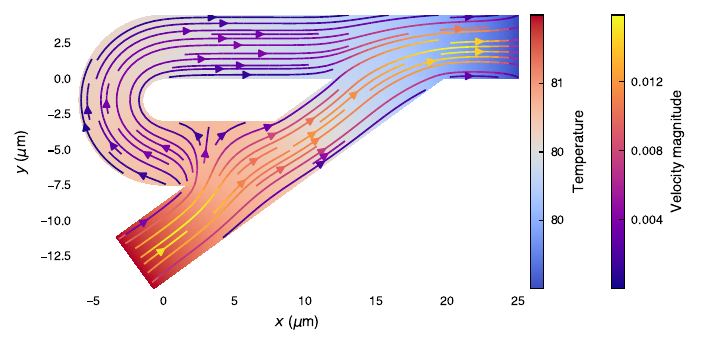}
        \caption{}
        \label{fig:field_left}
    \end{subfigure}
    \hfill
    \begin{subfigure}{0.48\textwidth}
        \centering
        \includegraphics[width=\textwidth]{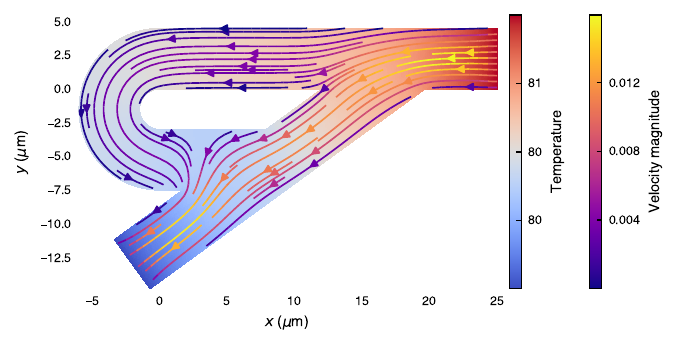}
        \caption{}
        \label{fig:field_right}
    \end{subfigure}
    \caption{Temperature field and velocity streamlines for left (a) and right (b) inlet with normal velocity 0.01 $\mu$m/ns.}
    \label{fig:fields}
\end{figure}

\begin{figure}[htbp]
    \centering
    \begin{subfigure}[t]{0.48\textwidth}
        \centering
        \includegraphics[width=\textwidth]{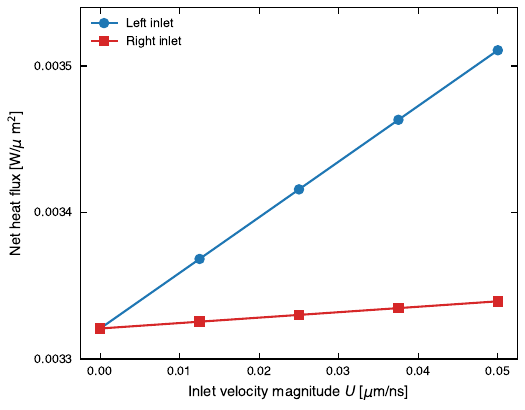}
        \caption{Heat flux sweep for the two cases: (1) non zero velocity at the left inlet and zero velocity at the right outlet and (2) non zero velocity at the right inlet and zero velocity at the left outlet.}
        \label{fig:heat_flux_sweep}
    \end{subfigure}
    \hfill
    \begin{subfigure}[t]{0.48\textwidth}
        \centering
        \includegraphics[width=\textwidth]{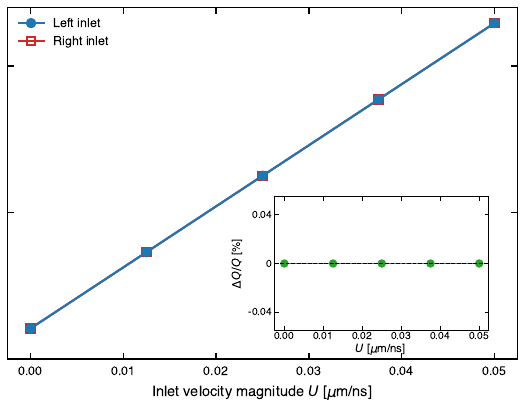}
        \caption{Heat flux for the (1) left and (2) right inlet, velocity of outlet is equal to the inlet velocity. The curves are visually indistinguishable; inset shows the relative residual $\Delta Q/Q = (Q_\mathrm{fwd}-Q_\mathrm{rev})/Q_\mathrm{fwd}$.}
        \label{fig:heat_flux_residual}
    \end{subfigure}
    \caption{Validation of left--right flux symmetry under reversed inlet/outlet driving conditions.}
    \label{fig:heat_flux}
\end{figure}

Returning to the analogy with (incompressible) classical fluids, conservation of mass would impose equal velocities at the inlet and outlet boundaries, precluding any rectification effect in the Stokes regime. In the phonon fluid, however, phonon number is not conserved due to inelastic (resistive) scattering processes, which relaxes this constraint and permits the rectification behavior described above. 

A direct test would be to repeat the calculations in Ref.~\cite{WuHu2026} with true interchanged thermal-reservoir boundary conditions, \(F_\lambda=C_\lambda\Theta\) on incoming modes at terminal 1 and \(F_\lambda=0\) on incoming modes at terminal 2, followed by the reversed problem with these incoming populations interchanged. The fixed-background linearized BTE then predicts equal resistance magnitudes, independent of hydrodynamic phonon transport, normal scattering, anisotropic mode redistribution, or asymmetric heat-flux streamlines. Therefore, if \(R_{\mathrm{f}}/R_{\mathrm{b}}\neq 1\) persists only under a gradient-weighted inlet source and absorbing-outlet protocol, it should be interpreted as a source--sink response rather than a two-terminal thermal-rectification ratio. True two-terminal rectification would require physics beyond the passive fixed-background linearized BTE, such as temperature-dependent scattering at finite \(\Delta T\), nonlinear boundary conditions, active elements, additional fixed terminals, or explicit time-reversal-symmetry breaking.

\end{document}